\documentclass[preprint,aps,nofootinbib]{revtex4-1}
\pdfoutput=1

\usepackage{amsmath}  
\usepackage{amsbsy,amssymb,amsfonts}
\usepackage{graphicx}
\usepackage{color}
\usepackage{epsf}
\usepackage[utf8x]{inputenc}
\usepackage{hyperref}
\usepackage{multirow}
\def\ie{{\em{i.e.}},}
\def\eg{e.g.\ }

\def\bravert{\egroup\,\vrule\,\bgroup}

{\catcode`\|=\active
  \gdef\Twoint#1{\left(\mathcode`\|"8000\let|\bravert {#1}\right)}}
{\catcode`\|=\active
  \gdef\Braket#1{\left<\mathcode`\|"8000\let|\bravert {#1}\right>}}

\newcommand{\beq}{\begin{equation}}
\newcommand{\eeq}{\end{equation}}
\newcommand{\beqa}{\begin{eqnarray}}
\newcommand{\eeqa}{\end{eqnarray}}
\newcommand{\bea}{\begin{array}}
\newcommand{\eea}{\end{array}}

\newcommand{\bef}{\begin{figure}}
\newcommand{\ef}{\end{figure}}
\newcommand{\bc}{\begin{center}}
\newcommand{\ec}{\end{center}}
\newcommand{\bt}{\begin{table}}
\newcommand{\et}{\end{table}}
\newcommand{\btb}{\begin{tabular}}
\newcommand{\etb}{\end{tabular}}

\newcommand{\au}{{\em a.u.}}

\begin{document}

\title {Model-independent determinations of the electron EDM and the role of diamagnetic atoms}

\vspace*{1cm}

\author{Timo Fleig}
\email{timo.fleig@irsamc.ups-tlse.fr}
\affiliation{Laboratoire de Chimie et Physique Quantiques,
             IRSAMC, Universit{\'e} Paul Sabatier Toulouse III,
             118 Route de Narbonne, 
             F-31062 Toulouse, France }
\author{Martin Jung}
\email{martin.jung@tum.de}
\affiliation{Excellence Cluster Universe \\
Technische Universit\"at M\"unchen\\ Boltzmannstr. 2, D-85748 Garching, Germany
\vspace*{1cm}
}
\begin{abstract}
 We perform model-independent analyses extracting limits for the electric dipole moment  of the electron  and the
P,T-odd scalar-pseudoscalar (S-PS) nucleon-electron coupling from the most recent measurements with atoms and molecules. The
analysis using paramagnetic systems, only, is improved substantially by the inclusion of the recent measurement on HfF${}^+$ ions,
but complicated by the fact that the corresponding constraints are largely aligned, owing to a general relation between the
coefficients for the two contributions.
Since this same relation does not hold in diamagnetic systems, it is possible to find atoms that provide essentially orthogonal
constraints to those from para\-magnetic ones. 
However, the coefficients are suppressed in closed-shell systems and enhancements of P,T-odd effects are
only prevalent in the presence of hyperfine interactions.
We formulate the hyperfine-induced time-reversal-symmetry breaking S-PS nucleon-electron interaction in general atoms in a mixed
perturbative and variational approach, based on electronic Dirac-wavefunctions including the effects of electron correlations. The
method is applied to the Hg atom, yielding the first direct calculation of the coefficient of the S-PS
nucleon-electron coupling in a diamagnetic system. This results in  additionally improved
model-independent limits for both the electron EDM and the nucleon-electron coupling from the global fit. Finally we employ this
fit to provide indirect limits for several paramagnetic systems under investigation.

\end{abstract}

\maketitle
\section{Introduction}
 Electric dipole moments (EDMs) provide a competitive means to search for new physics (NP), complementary to strategies like direct
searches at hadron colliders, but also to other indirect searches, for instance using flavour-changing processes.
The exceptional sensitivity  is due to the combination of experimental precision
with a tiny Standard Model (SM) background.\footnote{\emph{Strong CP violation} constitutes a potential exception to this statement;
however, the neutron EDM indicates a tiny value for this contribution as well.} Experimental tests for EDMs involve
typically rather complex systems like atoms or molecules. The discovery of a finite EDM in any of these systems would be a major
discovery, independent of its specific source.
However, reliably interpreting these measurements in terms of fundamental parameters of a given NP model requires precise knowledge of
their relations. 
These are established proceeding via a series of effective field theories, rendering a large part of the analysis model- and
system-independent, see \emph{e.g.} Refs.~\cite{ginges_flambaum2004,EDMsNP_PospelovRitz2005,Raidal:2008jk,fukuyama_review2012,
ramsey-musolf_review2_2013,Bsaisou:2014oka,Chupp:2017rkp} for recent reviews. The corresponding complex matrix elements on the atomic,
nuclear and QCD levels  often involve large uncertainties, which sometimes prohibit to fully exploit the experimental information, see
Refs.~\cite{MJung_review2014,ramsey-musolf_review2_2013} for recent detailed discussions.

This article presents a new method for the rigorous calculation of the coefficient of
the scalar-pseudoscalar nucleon-electron (S-PS-ne) interaction in diamagnetic systems. For this contribution so far only
rough estimates exist, due to the fact that it vanishes to leading order in the electromagnetic interaction, even in the
presence of an external electric field.
In this paper  we consider Mercury (Hg) which provides the strongest experimental limit on an EDM so far \cite{Heckel_Hg_PRL2016}.
The determination of this coefficient provides a competitive limit on the (NP-induced) strength of the corresponding interaction. 
It is also of special interest for the model-independent extraction of the electron EDM: in principle, paramagnetic systems can be
used to obtain both coefficients, taking into account potential cancellations~\cite{PhysRevA.85.029901,MJung_robustlimit2013};
however, a problem arises from the fact that all paramagnetic systems constrain a similar combination of these two 
contributions~\cite{PhysRevA.85.029901}.
Diamagnetic systems generally give independent constraints, thereby improving the model-independent extraction of both coefficients
significantly~\cite{MJung_robustlimit2013}. Our results can therefore be used to constrain different classes of NP models,
requiring less restrictive assumptions.

This article proceeds as follows: In the following section we present a method for the direct calculation of S-PS-ne enhancements in
closed-shell atoms and molecules. Section \ref{SEC:APPLICATION} describes its application to the Hg atom, and in section 
\ref{SEC:PHENOMENOLOGY} we investigate the phenomenological consequences of the present study. In the final
section we conclude and discuss the implications of our findings for future work.

\section{Theoretical Framework}
The calculation of the dominant contribution induced by the S-PS-ne interaction in diamagnetic systems requires the inclusion of
the hyperfine interaction on top of the corresponding calculation in paramagnetic systems, since its expectation value vanishes to
leading order in a closed-shell atom, due to a vanishing spin density near its nucleus 
\cite{schiff_nucEDM_1963,khriplovich_lamoreaux}. The nuclear current at the origin, corresponding to the magnetic moment of the
nucleus, polarizes the closed atomic shells and leads to non-zero values. In a traditional setup this would require a three-fold
expansion in the S-PS-ne interaction, the external electric field and the hyperfine interaction. Instead, we here start from a
0th-order electronic-structure problem
\begin{equation}
 \hat{H}^{(0)}\, \left| \psi^{(0)}_K \right> = \varepsilon_K^{(0)}\, \left| \psi^{(0)}_K \right>\,,
 \label{EQ:SEV_EQ}
\end{equation}
where $H^{(0)}$ is the atomic Dirac-Coulomb Hamiltonian \emph{including} the perturbation due
to a homogeneous external electric field ${\bf{E}_{\text{ext}}}$, with the nucleus placed
at the origin:
\begin{eqnarray}
 \nonumber
\hat{H}^{(0)} &:=& \hat{H}^{\text{Dirac-Coulomb}} + \hat{H}^{\text{Int-Dipole}} \\
&=& \sum\limits^N_j\, \left[ c\, \boldsymbol{\alpha}_j \cdot {\bf{p}}_j + \beta_j c^2 
 + \frac{Z}{r_j}{1\!\!1}_4 \right]
     + \sum\limits^N_{j,k>j}\, \frac{1}{r_{jk}}{1\!\!1}_4
        + \sum\limits_j\, {\bf{r}}_j \cdot {\bf{E}_{\text{ext}}}\, {1\!\!1}_4\,,
 \label{EQ:HAMILTONIAN}
\end{eqnarray}
where the indices $j,k$ run over $N$ electrons, $Z$ is the proton number ($N=Z$ for neutral atoms), and
$\boldsymbol{\alpha},\beta$ are standard Dirac matrices. 
We use atomic units (\au) throughout ($e = m_0 = \hbar = 1$).
Since we solve  Eq.~\eqref{EQ:SEV_EQ} variationally (\ie\ by diagonalization), the effect of the external electric field in
$\left| \psi^{(0)}_K \right>$ is taken into account to all orders in perturbation theory.
These states are technically electronic configuration interaction (CI) vectors \cite{knecht_luciparII}.

The first-order perturbed wavefunction due to the magnetic hyperfine interaction can be written as
\begin{equation}
 \label{EQ:HYPFIN_PERTSUM1}
\left| \psi^{(1)}_J \right> = \left| \psi^{(0)}_J \right>
                               + \sum\limits_{K \ne J}\, \frac{ \left< \psi^{(0)}_K \right| \hat{H}^{(1)}_{\text{HF}}
                                               \left| \psi^{(0)}_J \right> }
                                              { \varepsilon_J^{(0)} - \varepsilon_K^{(0)} }\,
                               \left| \psi^{(0)}_K \right>,
\end{equation}
where in practice the summation is carried out over a restricted set of CI vectors.
The perturbation sum in Eq. (\ref{EQ:HYPFIN_PERTSUM1}) will only be well-defined if $\left| \psi^{(0)}_J \right>$ is a 
non-degenerate state, which is the case for the electronic ground state of a closed-shell atom.

Since $\hat{H}^{(1)}_{\text{HF}}$ is a totally symmetric operator with respect to all valid symmetry operations of the
system including the external field (axial symmetry), the sum in Eq. (\ref{EQ:HYPFIN_PERTSUM1}) includes only states of the same
irreducible representation as the reference state $\left| \psi^{(0)}_J \right>$. The magnetic hyperfine Hamiltonian reads
\begin{equation}
\hat{H}_{\text{HF}} = - \frac{1}{2c\, m_p}\, \frac{\mu \boldsymbol{I}}{I}\,
       \cdot \sum\limits_{i=1}^n\,
       \frac{\boldsymbol{\alpha}_i \times \boldsymbol{r}_i}{r_i^3} \,,
 \label{EQ:HAM_HYPFIN}
\end{equation}
where $\mu = gI$ is the nuclear magnetic moment, $g$ the nuclear $g$-factor, $m_p$ the proton mass 
and $\boldsymbol{I}$ the nuclear spin.
The minus sign in Eq. (\ref{EQ:HAM_HYPFIN}) relates to the charge of an electron in \au\
The hyperfine Hamiltonian can also be written as $\hat{H}^{(1)}_{\text{HF}} = \boldsymbol{I}\, {\cal{A}}\,
\boldsymbol{J}$, where ${\cal{A}}$ is the rank 2 cartesian hyperfine interaction tensor and $\boldsymbol{J}$ is the 
total electronic angular momentum. It is, therefore, generally a sum of nine terms that due to 
$\mu := \left< I, M_I=I | \hat{\mu}_z | I, M_I=I \right>$ and $\boldsymbol{\mu} \propto \boldsymbol{I}$ reduces
to $\hat{H}^{(1)}_{\text{HF}} = I_z \left( {\cal{A}}_{zx} J_x + {\cal{A}}_{zy} J_y + {\cal{A}}_{zz} J_z \right)$. 
The required matrix elements are defined as follows:
\begin{equation}
\left(A_{zk}\right)_{MN} = - \frac{\mu[\mu_N]}{2c I m_p}\, \sum\limits_{i=1}^n\, 
 \left< \psi^{(0)}_M \right| \left( \frac{\boldsymbol{\alpha}_i \times \boldsymbol{r}_i}{r_i^3} \right)_k
 \left| \psi^{(0)}_N \right>\,,
\end{equation}
where $k$ is a cartesian component and the nuclear magnetic moment enters in units of the nuclear magneton 
$\mu_N = \frac{1}{2cm_p}$ (in \au).

For evaluating the S-PS-ne enhancement in the atom we use the effective Hamiltonian operator \cite{Flambaum_Khriplovich1985}
\begin{equation}
 \hat{H}_{\text{S-PS-ne}}^{(1)}(S) = \imath \frac{G_F}{\sqrt{2}}\, AC_S\, 
                                \sum\limits_e\,\gamma^0_e\, \gamma^5_e\, \rho({\bf{r}}_e)\,,
 \label{EQ:HAM_NE-SPS}
\end{equation}
where $G_F$ is the Fermi constant, $A$ the nucleon number, $C_S$ the dimensionless S-PS-ne coupling constant, $\rho$ the normalized
nuclear charge density, and $\gamma^{\mu}$ are standard Dirac matrices.
Given the smallness of this interaction, even compared to the hyperfine interaction, higher-order perturbative corrections are
clearly negligible. Given, furthermore, the CP-conserving nature of the hyperfine interaction, the energy shift of a given
atomic state indicating CP violation can to leading order be written as
\begin{equation}
    \left(\Delta \varepsilon\right)_J = 
    \frac{1}{\langle \psi^{(1)}_J|\psi^{(1)}_J\rangle}\left<
    \hat{H}_{\text{S-PS-ne}}^{(1)}\right>_{\psi^{(1)}_J}.
\end{equation}
The atomic EDM in terms of the S-PS-ne interaction is a function of the polarizing external electric field 
$E_{\text{ext}}$, and so
\begin{align}
 d_a = 
 -\lim\limits_{E_{\text{ext}} \rightarrow 0}\, \left[\frac{\partial (\Delta \varepsilon)}{\partial E_{\text{ext}}} \right]
 \approx - A C_S \frac{G_F}{\sqrt{2}}\, \frac{\left< \imath \sum\limits_e\, \gamma^0_e\, \gamma^5_e\, \rho({\bf{r}}_e) 
    \right>_{\psi^{(1)}(E_{\text{ext}})}}{E_{\text{ext}}\,\langle \psi^{(1)}|\psi^{(1)}\rangle}\equiv \alpha_{C_S}C_S\,,
  \label{EQ:ATOMIC_EDM}
\end{align}
where the approximation holds in the linear regime
which is assured by external fields chosen significantly smaller than the internal ones and we have introduced $\alpha_{C_S}$,
the atomic S-PS-ne enhancement factor.
In the present case $E_{\text{ext}}$(Hg) $= 0.00024$ \au\
This leads to shifts of the energies $\varepsilon_K^{(0)}$ (see Eq. (\ref{EQ:SEV_EQ})) on the order of
$10^{-6}$ \au\ for Hg. CI vectors are consequently
optimized such that the energies $\varepsilon_K^{(0)}$ are converged to at least $10^{-9}$ \au\

We now focus on the evaluation of the normalized expectation value, part of the expression on the right-hand side of Eq.
(\ref{EQ:ATOMIC_EDM}),
\begin{eqnarray}
\label{EQ:WS_PERTHF} 
&& \frac{1}{\left< \psi^{(1)}_J \right| \left. \psi^{(1)}_J \right>}\,
          \left< \psi^{(1)}_J \right| \imath\, \sum\limits_e\,\gamma^0_e\, \gamma^5_e\, \rho({\bf{r}}_e)
          \left| \psi^{(1)}_J \right>=\nonumber\\
&&\frac{1}{\left< \psi^{(1)}_J \right| \left. \psi^{(1)}_J \right>}\,
             {\left. \rule{0.0cm}{1.0cm} \right[ }
     \sum\limits_{K \ne J}\, \frac{ \left< \psi^{(0)}_K \right| \hat{H}^{(1)}_{\text{HF}}
               \left| \psi^{(0)}_J \right> }
         { \varepsilon_J^{(0)} - \varepsilon_K^{(0)} }\,
     \left< \psi^{(0)}_J \right| \imath\, \sum\limits_e\,\gamma^0_e\, \gamma^5_e\, \rho({\bf{r}}_e)
          \left| \psi^{(0)}_K \right> \\
&&   \left.\hspace*{2.5cm} + \sum\limits_{K \ne J}\, \frac{ \left< \psi^{(0)}_J \right| \hat{H}^{(1)}_{\text{HF}}
             \left| \psi^{(0)}_K \right> }
             { \varepsilon_J^{(0)} - \varepsilon_K^{(0)} }\,
         \left< \psi^{(0)}_K \right| \imath\, \sum\limits_e\,\gamma^0_e\, \gamma^5_e\, \rho({\bf{r}}_e)
            \left| \psi^{(0)}_J \right> 
       \right]\,,
 \nonumber
\end{eqnarray}
up to higher-order terms in the hyperfine interaction,
where we used the hyperfine-perturbed wavefunction from Eq. (\ref{EQ:HYPFIN_PERTSUM1}).
The leading term in this equation (for open-shell atoms) vanishes for closed-shell atoms, and is omitted.
This conclusion has also been tested numerically in the present work. Transition matrix elements of the type
$\left< \psi^{(0)}_K \right| \hat{H}^{(1)}_{\text{HF}} \left| \psi^{(0)}_J \right>$ and
$\left< \psi^{(0)}_K \right| \imath\, \sum\limits_e\,\gamma^0_e\, \gamma^5_e\, \rho({\bf{r}}_e)
        \left| \psi^{(0)}_J \right>$, required for evaluating these two terms, can be readily 
made available using the developed methodology in Refs. \cite{Fleig2014,ThF+_NJP_2015}. The practical problem 
is then to provide a sufficient set of CI states for the perturbation sum. The final expression for
evaluating the S-PS-ne enhancement is, therefore,
\begin{equation}
 \alpha_{C_S}(\psi_J) = 
 \frac{-A \frac{G_F}{\sqrt{2}}}{E_{\text{ext}}\, \left< \psi^{(1)}_J \right| \left. \psi^{(1)}_J \right>}\,
              \left[
          \sum\limits_{K \ne J}\, \frac{ \left< \psi^{(0)}_K \right| \hat{H}^{(1)}_{\text{HF}}
           \left| \psi^{(0)}_J \right> }
           { \varepsilon_J^{(0)} - \varepsilon_K^{(0)} }\,
        \left< \psi^{(0)}_J \right| \imath\, \sum\limits_e\,\gamma^0_e\, \gamma^5_e\, \rho({\bf{r}}_e)
	\left| \psi^{(0)}_K \right> + h.c. \right]
\end{equation}
For convenience, we use in the following also the S-PS-ne enhancement $S$ (in analogy to the electron EDM
enhancement $R$ and not to be confused with the nuclear
Schiff moment, also denoted $S$ in the literature), defined as 
\begin{align}
 S := \frac{d_a}{A C_S \frac{G_F}{\sqrt{2}}}=\frac{\alpha_{C_S}}{A\frac{G_F}{\sqrt{2}}}
 \approx - \frac{\left< \imath \sum\limits_e\, \gamma^0_e\, \gamma^5_e\, \rho({\bf{r}}_e) 
    \right>_{\psi^{(1)}(E_{\text{ext}})}}{E_{\text{ext}}\,\langle \psi^{(1)}|\psi^{(1)}\rangle}.
  \label{EQ:SRATIO}
\end{align}

In order to facilitate comparison with the literature, we note that the states $\left| \psi^{(0)}_K \right>$ can be considered as
wavefunctions perturbed to infinite order by ${\bf{E}}$, and so the expression in Eq. (\ref{EQ:WS_PERTHF}) contains terms of third order of the type
\begin{equation}
    \sum\limits_{K,N \ne J} 
    \frac{ \left< \psi^{(0)}_J \right| \sum\limits_i\, \hat{r}_z(i) \left| \psi^{(0)}_N \right> E_z\,
    \left< \psi^{(0)}_K \right| \hat{H}^{(1)}_{\text{HF}} \left| \psi^{(0)}_J \right> }
      { \left( \varepsilon_J^{(0)} - \varepsilon_N^{(0)} \right) 
        \left( \varepsilon_J^{(0)} - \varepsilon_K^{(0)} \right) 
      } \left< \psi^{(0)}_N \right| \imath\, \sum\limits_e\,\gamma^0_e\, \gamma^5_e\, \rho({\bf{r}}_e)
        \left| \psi^{(0)}_K \right>,
 \label{EQ:WS_PERTE_PERTHF3}
\end{equation}
plus higher-order contributions in $\mathbf E$, where $\left| \psi^{(0)}_N \right>$ is now an unperturbed eigenstate of the plain
atomic Dirac-Coulomb Hamiltonian {\it{without}} external electric field.
The terms in Eq. (\ref{EQ:WS_PERTE_PERTHF3}) are just the equivalent of the electron EDM contribution via
magnetic hyperfine interaction to an atomic EDM, as given by Flambaum and Khriplovich in reference 
\cite{Flambaum_Khriplovich1985}, Eq. (17). These third-order terms, declared important but left untreated
in reference \cite{dzuba_flambaum_PRA2009}, are taken into account  
in the present approach. Moreover, the higher-order contributions in $\mathbf E$ are included automatically in the present approach.

 \section{ne-SPS enhancement in atomic mercury}
  \label{SEC:APPLICATION}

For our zeroth-order atomic wavefunctions the quantum
number $M_J$, corresponding to the projection of the total angular momentum onto the axis defined by the external 
electric field, is an exact quantum number and characterizes an irreducible representation of
the axial double point group.
Since the external perturbation is small, the quantum number $J$ is still approximately valid and we denote CI states in the
approximate Russell-Saunders picture as ${^ML_{J,M_J}}$, where $M$ is the spin multiplicity.
The S-PS-ne interaction Hamiltonian in Eq. (\ref{EQ:HAM_NE-SPS}) is rotationally invariant;
as a consequence, 
$\left< M_J |\hat{H}_{\text{S-PS-ne}} | M_J' \right> = 0$ for  $M_J \ne M_J'$, which
reduces the perturbation sum in Eq. (\ref{EQ:WS_PERTHF}) to states from the irreducible
representation $M_J = 0$, a computational advantage which we exploit.

Applying the framework developed in the last section to Mercury,
a consistent finding in all our calculations is that among the $35$ energetically
lowest-lying excited states of symmetry $M_J = 0$ only three states contribute sizably
to the perturbation sum Eq. (\ref{EQ:WS_PERTHF}) determining $\alpha_{C_S}$, namely
$\psi^{(0)}_K \in \{ {^3P_{0,M_J=0}} (5d^{10}6s6p),$ ${^3S_{1,M_J=0}} (5d^{10}6s7s),
{^3P_{0,M_J=0}} (5d^{10}6s7p) \}$. This finding can be 
understood qualitatively analyzing
the product of matrix elements in Eq. (\ref{EQ:WS_PERTHF}): For contributions of the type
\begin{displaymath}
 \left< {^3P_{0,M_J=0}} \right| \hat{H}^{(1)}_{\text{HF}} \left| {^1S_{0,M_J=0}} \right>
 \left< {^1S_{0,M_J=0}} \right| \imath\, \sum\limits_e\,\gamma^0_e\, \gamma^5_e\, \rho({\bf{r}}_e)
 \left| {^3P_{0,M_J=0}} \right>
\end{displaymath}
the off-diagonal S-PS-ne matrix element is 
large due to the parity-odd excitation $6s \rightarrow np$ characterizing the excited state, and the off-diagonal hyperfine matrix element is
non-negligible due to $sp$-mixing {\it{via}} the external electric field. For the other leading type of contribution,
\begin{displaymath}
 \left< {^3S_{1,M_J=0}} \right| \hat{H}^{(1)}_{\text{HF}} \left| {^1S_{0,M_J=0}} \right>
 \left< {^1S_{0,M_J=0}} \right| \imath\, \sum\limits_e\,\gamma^0_e\, \gamma^5_e\, \rho({\bf{r}}_e)
 \left| {^3S_{1,M_J=0}} \right>\,,
\end{displaymath}
the off-diagonal S-PS-ne matrix element is now two orders of magnitude smaller than in the above case -- for
obvious reasons related to symmetry --, but the off-diagonal hyperfine matrix element becomes almost
three orders of magnitude larger than for the previous mechanism.
This is explained by the fact that the excited state ${^3S_1}$ exhibits a non-vanishing spin-density near the nucleus.

Results from many-body models of different sophistication are compiled in Table~\ref{TAB:199Hg:TZ_AEHYEN_SDT12_SD12_150au}. The S-PS-ne
enhancement is largely converged when at least the 12 lowest-lying $M_J = 0$ states are included in the perturbation sum, since
then the three main contributors are covered. 
%
\begin{table}[th]
\centering{
 \begin{tabular}{l@{\hspace{4mm}}l@{\hspace{4mm}}c@{\hspace{4mm}}c@{\hspace{4mm}}c}\hline\hline
 Basis/cutoff 
 & \# of CI states $M_J=0$/Model/X & $\frac{\rm Mean dev.}{\%}$ 
 &  $\frac{S}{10^{-2}a.u.}$ & $\frac{\alpha_{C_S}}{10^{-22}\, e{\rm cm}}$ \\ \hline
  DZ/150 a.u.& $4$/M12/6p7s7p6d5f8p8s7d        &       &  $-3.3$                 & $-5.4$   \\
  DZ/150 a.u.& $16$/M12/6p7s7p6d5f8p8s7d       &        & $-2.3$                 & $-3.8$   \\
  TZ/50 a.u. & $12$/M12/6p7s7p                 & $6.1$ &  $-2.1$                 & $-3.5$   \\
  TZ/50 a.u. & $12$/M20/6p7s7p                 &       &  $-2.1 $                 & $-3.5 $   \\
  TZ/50 a.u. & $12$/M12/6p7s7p6d8p8s           & $5.4$ &  $-2.2 $                 & $-3.7 $   \\
  TZ/50 a.u. & $29$/M12/6p7s7p6d8p8s9p9s10p10s & $6.2$ &  $-2.22$                 & $-3.67$   \\ \hline\hline
 \end{tabular}
\caption{\label{TAB:199Hg:TZ_AEHYEN_SDT12_SD12_150au}
    S-PS-ne interaction  
    ratio $S$ for the ${^1S}_0$ ground state 
    of the {$^{199}$Hg} isotope, $I=1/2$, 
    $\mu({^{199}{\text{Hg}}}) = +0.5058855$ \cite{stone_INDC2015},
    $E_{\text{Ext}} = 0.00024$ \au; 
    CI models M12: 12 electrons correlated, Single, Double and Triple excitations from occupied space into X, 
    Single and Double excitations into the remaining virtual space (SDT12-X-SD12); M20: S8-SDT12-X-SD20. 
    DZ and TZ denote Dyall's Gaussian atomic basis sets \cite{dyall_basis_2004,dyall_gomes_basis_2010} 
    including 1f,1g valence- and core-correlating exponents (DZ) and 
    2f,4g,1h valence- and core-correlating and valence-polarizing exponents (TZ), resulting in a total of
    24s,19p,12d,8f,1g for DZ and 30s,24p,15d,11f,4g,1h functions for TZ. The mean deviation
    concerns the difference of the calculated excited-state energies from experiment \cite{NIST_Hg}.
    The Hg nucleus is described by a Gaussian charge distribution \cite{visscher_dyall_nucdist}
    with exponent $\zeta = 1.4011788914 \times 10^8$.
         }
}
\end{table}

It is furthermore important  that the extent of the
active spinor space is sufficient, as can be seen from the results for different values of $X$, the parameter defining the
atomic functions constituting the space into which triple excitations are allowed. The remaining virtual spinors up to the cutoff threshold
are allowed to be up to doubly occupied, in order to include dynamic electron correlation effects for
all states described to lowest order by the structure of the active space. Correlation effects between
$5s,5p$ and valence electrons are tested through the model including $20$ electrons and are seen to be
small.

For the purpose of estimating the contribution from higher-lying excited states we use a larger basis set,
denoted QZ and consisting of 34s,30p,19d,13f,4g,2h functions. Due to computational demand the model M12 is
limited to X-SDT12 with X set to the value 7p7s8p9p8s10p9s with reference to Table 
\ref{TAB:199Hg:TZ_AEHYEN_SDT12_SD12_150au}.
This means that correlation effects are largely neglected for a large set of small contributions, $\approx
100$ states with $M_J = 0$. We observe that only two notable contributions occur, and only in the energetically 
lower half, indicating that the contributions as expected fall off as energy and principal
quantum number of the involved states increase. With the resulting enhancement correction $\Delta S$(QZ),
where $S$ is defined in Eq. (\ref{EQ:SRATIO}), our final value is obtained as follows:
\begin{equation}
 \label{EQ:FINAL_S}
 S{\text{(TZ)}} + \Delta S{\text{(QZ)}} = (-2.22 + 0.53)\times 10^{-2} a.u. = -1.69\times 10^{-2}\, a.u.
\end{equation}
The uncertainty of this value is estimated by \emph{linearly} adding the errors from the energy denominator 
($6.2$\%, ``mean deviation''
in Table~\ref{TAB:199Hg:TZ_AEHYEN_SDT12_SD12_150au}), and uncertainties from atomic basis set ($3.5$\%), outer-core
correlations ($1.5$\%), and higher excitation ranks ($5$\%, estimated from comparable previous calculations of S-PS-ne 
enhancements, see Refs.~\cite{Denis-Fleig_ThO_JCP2016,fleig:PRA2016}). To this uncertainty of $16$\% on the base value
$S{\text{(TZ)}}$ we add an uncertainty of $30$\% times the relative weight ($0.24$) of the correction $\Delta S$(QZ), \ie
$7.2$\%, resulting in a total uncertainty of $23$\% for $\alpha_{C_S}$, which we consider very conservative. 
Note that adding the individual terms in quadrature, as commonly done in the literature, would result in an uncertainty
of $11\%$.
From these considerations, we finally obtain from Eq.~\eqref{EQ:FINAL_S} the  S-PS-ne interaction constant
\begin{equation}\label{eq::alphaCS}
 \alpha_{C_S} = -2.8(6)\times 10^{-22}\, e\,{\rm cm}\,. 
\end{equation}

An indirect determination of $\alpha_{C_S}$ is obtained {\it{via}} the coefficient of the (P,T)-odd tensor interaction,
using the phenomenological relation \cite{Flambaum_Khriplovich1985,Kozlov:1988qn,khriplovich_lamoreaux}
\begin{align}\label{eq::CSCT}
\frac{\langle\boldsymbol{\sigma}\rangle\cdot \mathbf I}{I}\alpha_{C_S} = 5.3\times
10^{-4}(1+0.3\,Z^2\alpha^2)A^{2/3}\mu_A \alpha_{C_T}\,,
\end{align}
where $\langle \boldsymbol{\sigma}\rangle C_T \equiv \langle\sum_{N=n,p}C_T^{N}\boldsymbol{\sigma}_N\rangle$ ($\langle\ldots\rangle$
denoting the expectation value over a nuclear state with spin $\mathbf I$),
$\mu_A$ denotes the magnetic moment of the atom's nucleus (in units of the nuclear magneton), and the coefficients $C_T^N$ parametrize
the tensorial (P,T)-odd electron-nucleon interaction,
\begin{equation}
\mathcal H_T = \frac{iG_F}{\sqrt{2}}\sum_{N=n,p}C_T^N(\bar N\sigma_{\mu\nu}\gamma_5 N)(\bar e \sigma_{\mu\nu}e)\,.
\end{equation}
To further facilitate the comparison with other works, we note that the coefficient of the tensor interaction is 
typically parametrized {\it{via}}
$\mathbf d_A=10^{-20}C_{C_T}\langle\boldsymbol\sigma\rangle C_T\, e\,{\rm cm}$, implying 
\begin{equation}
\alpha_{C_T}=10^{-20}C_{C_T}\frac{\langle\boldsymbol{\sigma}\rangle\cdot \mathbf I}{I}e\,{\rm cm}\,.
\label{EQ:ACT_CCT}
\end{equation}
The comparison is shown in Table~\ref{TAB:LITERATURE}. 
\begin{table}[th]
\centering{
\begin{tabular}{l c c c}\hline\hline
Method  & Ref.  & $C_{C_T}$  & $\alpha_{C_S}/(10^{-22}e\,{\rm cm})$\\\hline
RPA      & \cite{mar-pendrill-_PRL1985}     & -6.0  & (-6.0)\\
MCDHF  & \cite{Radziute:2015apa}            & -4.8  & (-4.8)\\
CI+MBPT  & \cite{dzuba_flambaum_PRA2009}    & -5.1  & (-5.1)\\
PRCC  & \cite{Latha:2009nq}                 & -4.3  & (-4.3)\\
CCSD${}^{(\infty)}$ & \cite{Yamanaka:2017mef} & -3.4  & (-3.4)\\
${\rm CCSD_pT(+)}$  & \cite{Singh:2014jca}  & -4.0  & (-4.0)\\
${\rm CCSD_pT(+)}$  & \cite{Sahoo:2016zvr}  & -3.2  & (-3.2)\\

${\rm NCCSD}$  & \cite{2018arXiv180107045S}  & -3.3  & (-3.3)\\\hline
Chupp {\it{et al.}} (est.)  & \cite{Chupp_Ramsey_Global2015,Chupp:2017rkp} & & $(-5.9)$ \\
 Engel {\it{et al.}} (est.)  & \cite{ramsey-musolf_review2_2013} & & $(-8.1)$ \\\hline
\multicolumn{2}{l}{This work}              & (-2.8) &  $\boldsymbol{-2.8}$ \\ \hline\hline
 \end{tabular}\hfill
\caption{\label{TAB:LITERATURE}
Comparison of the direct calculation presented here with previous calculations of $\alpha_{C_S}$, using calculations of $\alpha_{C_T}$
and the phenomenological relation Eq.~\eqref{eq::CSCT} (indicated by parentheses around the result, $\mu_{Hg}=0.506$).
The literature values are ordered as to increasing sophistication of the treatment of dynamic electron correlation.
Numerically the conversion factor for Mercury reads $\alpha_{C_S}^{\rm Hg}=10^{-2}\alpha_{C_T}^{\rm Hg}/(\langle\boldsymbol
\sigma\rangle\cdot\mathbf  I/I)$, and a simple shell model for the nucleus is used, yielding $\langle\boldsymbol
\sigma\rangle\cdot\mathbf I/I=-1/3$. }
}
\end{table}
We note that effects of interelectron correlations reduce $C_{C_T}$ by about a factor of $1/2$. Due to relations
(\ref{eq::CSCT}) and (\ref{EQ:ACT_CCT}) these effects are expected to be qualitatively similar for the coefficient $\alpha_{C_S}$.  
In our result from the direct calculation electron correlation effects among the outermost $20$
electrons of the Hg atom have been taken into consideration. There are two main sources for a potential difference between our
value and the Coupled Cluster (CC) results {\it{via}} the phenomenological relation: 1) Our correlation model differs
from the correlation models used in the CC calculations. 2) The phenomenological relation employs a uniform
nuclear charge density whereas in our calculations a more realistic Gaussian charge distribution is used (see 
Table~\ref{TAB:199Hg:TZ_AEHYEN_SDT12_SD12_150au}) \cite{Visscher_Dyall_nuclcha}. 
Since correlation effects tend to reduce the absolute value of $\alpha_{C_S}$ and our value is already
about $15$\% below the CC results, it is reasonable to assume that no major correlation effects have been missed in our final
computational model. The present difference is furthermore within the expected precision of this relation.

\section{Phenomenological consequences}
 \label{SEC:PHENOMENOLOGY}

In order to explore the phenomenological consequences of our results, we follow two different strategies:
(i) The common method to limit the corresponding Wilson coefficients assuming the absence of cancellations, \emph{i.e.} setting
  all other contributions to zero.
(ii) Limiting \emph{both} $C_S$ and the electron EDM $d_e$ model-independently, \emph{i.e.} allowing for cancellations between the
two.
This is achieved by combining information from the Mercury system with that of paramagnetic ones, following
Ref.~\cite{MJung_robustlimit2013},  using the experimental results in
Table~\ref{tab::expinputs}.  The key point in this strategy is that Mercury constrains a linear combination of $d_e$ and $C_S$ that is
approximately orthogonal to the one constrained from paramagnetic systems, specifically ThO. This observation can be used
to constrain $C_S$ \emph{and} $d_e$, following a three-step argument:
%
\begin{table}[th]
\centering{
\begin{tabular}{c c c}\hline\hline
Molecule  &  $\omega_{\rm exp}/{\rm (mrad/s)}$ & Refs.\\\hline 
HfF$^+$ & $0.3\pm 2.7\pm0.6$\footnote{Adapted to match the conventions used here.} & \cite{Cairncross:2017fip}\\
ThO  & $2.6\pm4.8\pm3.2$ & \cite{ACME_ThO_eEDM_science2014,Baron:2016obh}\\
YbF  &  $5.3\pm 12.6\pm 3.8$ & \cite{hudson_hinds_YbF2011,Kara:2012ay}\\\hline
Atom  & $d_A/(e\,{\rm cm})$ & Refs.\\\hline
Tl    & $-(4.0\pm 4.3)\times 10^{-25}$ & \cite{regan_demille_2002}\\
Hg    & $(2.20\pm 2.75\pm1.48)\times 10^{-30}$ & \cite{Heckel_Hg_PRL2016}\\\hline\hline
\end{tabular}
\caption{\label{tab::expinputs} Experimental limits for the systems entering the global fit.} }
\end{table}
\begin{enumerate}
\item The EDMs of paramagnetic systems are to good approximation dominated by contributions from $d_e$ and $C_S$
\cite{sandars_PL1965,Sandars:1966xx,Flambaum:1976vg}.\footnote{Strictly speaking also contributions from the Schiff moment and in some
cases the magnetic quadrupole moment of the nucleus in paramagnetic systems could 
cancel these enhanced contributions. Given the large enhancement
of the latter by $Z^3\sim 10^5$, this would however imply huge contributions in other systems, which are at least as severely
constrained. However, formally a chain of cancellations in all constrained systems remains a possibility, due to the large number of
potential sources.} While $C_S$ depends in general on the system under consideration, the combination that enters heavy
atoms and molecules is to good approximation universal \cite{MJung_robustlimit2013}. $C_S$ cannot be neglected model-independently:
while  NP models exist where the electron EDM clearly gives the leading contribution, this is not true in
general.
In Two-Higgs-Doublet models (2HDMs) for instance, the dominating Barr-Zee diagram for the electron EDM avoids a second small mass factor
in addition to $m_e$, but as a two-loop diagram competes with a tree contribution to the S-PS-ne coupling that is suppressed by a
light-quark mass and contains additional small factors like gauge couplings \cite{MJung_review2014}. 
Schematically, we have  $m_{u,d,s}\times\,{\mbox{tree}}$ vs. $m_t \times\, {\mbox{two-loop}} \sim m_t/(16
\pi^2)^2$.
\item Both contributions can in principle easily be taken into account, once two experiments with comparable sensitivity are
    available. The problem is that most of the constraints from paramagnetic systems are essentially parallel, so that
    typically fine-tuned solutions exist, where electron EDM and S-PS-ne contributions 
    \emph{both} 
    oversaturate the experimental limit, but
    cancel to large extent in the measured observables. This leads to a situation where the model-independent approach yields a
    limit on the electron EDM that is about a factor of 10 weaker than the naive limit obtained when setting the S-PS-ne coupling to zero. This
    situation can be resolved by measurements on systems
    with different slopes, for example with relatively light atoms like Rb and very heavy ones like Fr. The recent measurement
    \cite{Cairncross:2017fip} already improves the situation significantly, as shown below.
\item In diamagnetic systems, there are many contributions to a potential EDM; assuming the presence of only electron EDM and S-PS-ne contributions
here is clearly not a good description of, {\eg}, the Mercury EDM. However, the different hierarchy in this case can be used to
    turn the argument around: In diamagnetic systems both contributions are \emph{not} enhanced, but strongly suppressed, because 
    they yield  a non-vanishing contribution only in combination with the hyperfine splitting. 
    The sensitivity of Mercury to the electron EDM is about $3 \times 10^8$ weaker than in ThO. The sensitivity to other contributions,
    like quark (C)EDMs, the theta term, and even tensor electron-nucleon couplings is much higher. This is why it is 
    conservative to assume
    that these  -- often neglected -- contributions saturate the experimental limit.
\end{enumerate}
The conditions that have to be met for the resulting limit to be invalid are consequently very specific: 
\begin{itemize}
\item The individual electron EDM and S-PS-ne contributions to the relevant paramagnetic systems would have to be larger than the
experimental limits, but cancel in all of them sufficiently well.
\item The electron EDM and S-PS-ne contributions to Hg would also have to be larger than the experimental limit, despite the
massively different sensitivity.
\item Since in the latter case a cancellation between the two contributions in Hg is not possible simultaneously with the paramagnetic
      systems, other contributions, that are each individually expected to be much larger than those from the electron EDM or S-PS-ne
      couplings, would have to combine in such a way that the net effect on the Hg EDM is again smaller than the experimental
      limit.
\end{itemize}
It is not impossible that all these things happen simultaneously, but since several cancellations on very different levels and in
very different systems are necessary, we consider the limit resulting from our procedure conservative. Assumptions are made only on a
subleading level, while in the literature it is very common to make them at the leading level, \emph{i.e.} simply neglecting the S-PS-ne
coupling. For convenience we provide below also the results without this assumption, \emph{i.e.} when using the data from paramagnetic
systems, only.

Note that the calculation presented here will remain useful even if the procedure outlined above should become unnecessary because
of measurements in paramagnetic systems providing sufficiently precise and non-parallel constraints. Ultimately the goal should be a
global analysis separating as many sources for EDMs as possible, see Ref.~\cite{Chupp_Ramsey_Global2015} for a first attempt. Should
both $d_e$ and $C_S$ be determined/limited from paramagnetic systems alone, the impact of the Mercury measurement on the remaining
sources will increase, given a sufficiently precise determination of the corresponding coefficients. 

Starting with strategy (i), \emph{i.e.} assuming $C_S$ to give the only contribution to the Mercury EDM, we obtain from
Ref.~\cite{Heckel_Hg_PRL2016} and  Eq.~\eqref{eq::alphaCS}
\begin{equation}
C_S = -\left(0.8^{+1.5}_{-1.2}\right)\times 10^{-8},\quad\mbox{or}\quad |C_S|\leq 3.2\times 10^{-8}\,(95\%~{\mbox{CL}})\quad
(C_S\mbox{ only).}
\end{equation}
This value is significantly larger than the one given in Ref.~\cite{Heckel_Hg_PRL2016}, for two reasons: Heckel
et al. used an indirectly obtained value for $\alpha_{C_S}$ \cite{mar-pendrill-_PRL1985}, where moreover
electron correlation effects have largely been neglected, which is much larger than our result on the absolute (and also larger
than newer indirectly obtained results), and presumably used only the central value of that result. It is also significantly larger
than the values obtained from ThO ($|C_S|\leq 0.7\times 10^{-8}\,(95\%~{\mbox{CL}})$) and HfF$^+$ ($|C_S|\leq 1.8\times
10^{-8}\,(95\%~{\mbox{CL}})$); However, as we will see below, the Hg result nevertheless improves the global fit significantly.

\begin{figure}[t]
\includegraphics[width=10.cm]{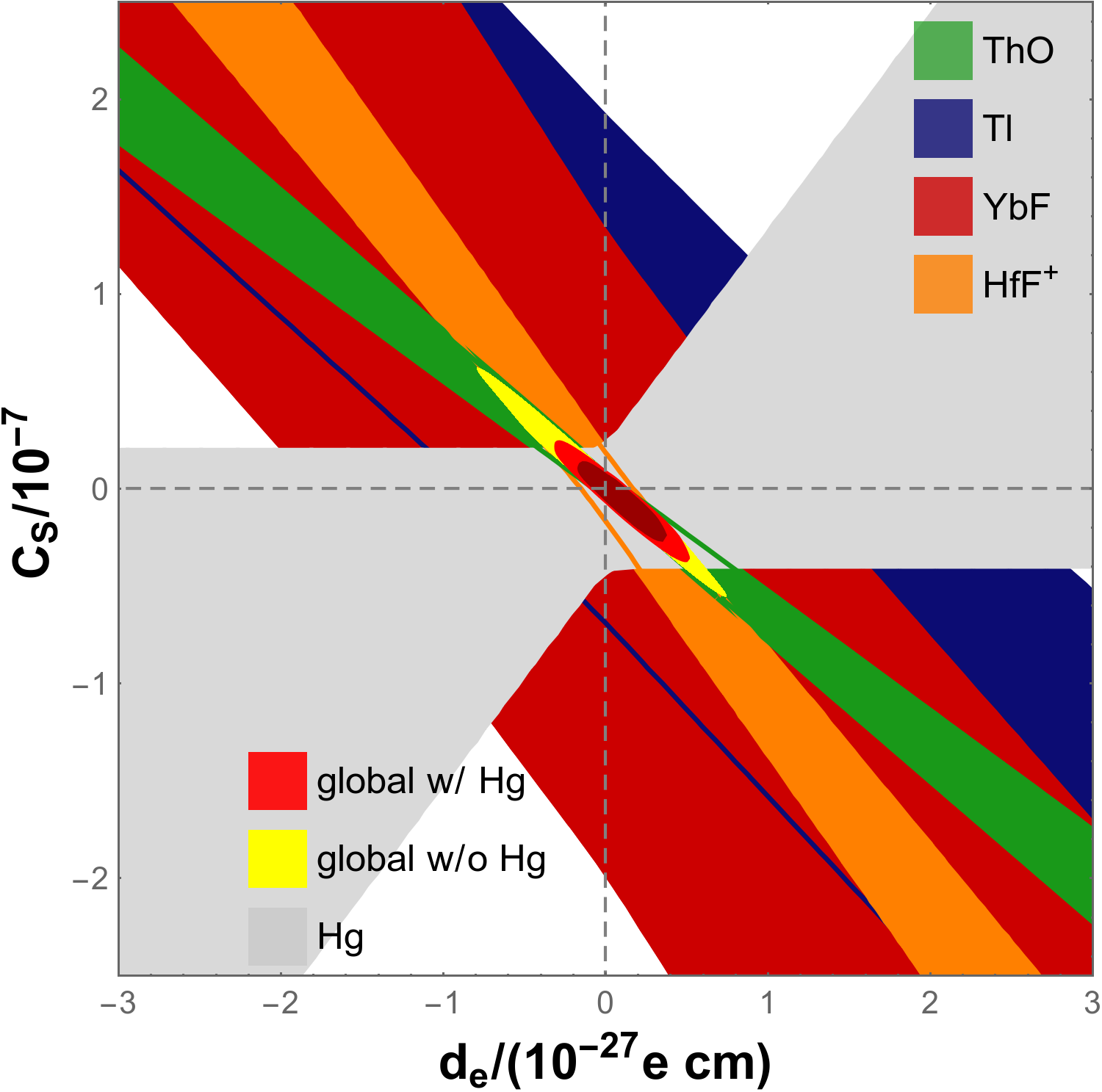}
\caption{\label{fig:fitdeCS} Fit to the available data from paramagnetic systems plus the constraint from Mercury, using the result
presented in this work. The bands from the individual constraints as well as the global fit without Mercury correspond to $95\%$~CL,
the global fit with Mercury to $68\%$ and $95\%$~CL. These bands include both experimental and theoretical uncertainties.
Individual constraints have 1 effective degree of freedom, the global fits 2.}
\end{figure}

We perform  global fits to the available data in Table~\ref{tab::expinputs},
using the theoretical inputs given in Table~\ref{tab::fitcoefficients}.  The  molecular measurements are typically expressed
in terms of the angular frequency $\omega_M$, which can for our purposes be written as 
\begin{align}\label{eq::omegaM0}
\omega_M &= \left(-1.52\, {\rm{sgn}}(\Omega)\frac{E_{\rm
eff}}{\rm{GV/cm}}\frac{d_e}{10^{-27}e\,{\rm
cm}}+2\pi\,10^6\,\Omega\frac{A_M}{Z_M}\frac{W_S}{\rm{kHz}}C_S\right)\langle\hat n\cdot \hat z\rangle\frac{\rm{mrad}}{\rm s}\\
&\equiv \alpha_{d_e}^M d_e+\alpha_{C_S}^M C_S\,,\label{eq::omegaM}
\end{align}
where $E_{\rm eff}$ the effective electric field, $\Omega = \left< {\bf{J}}_e \cdot {\bf{n}} \right>$ is the 
projection of the total electronic angular momentum ${\bf{J}}_e$ on the molecule-fixed internuclear axis ${\bf{n}}$, 
$\hat z$ is the laboratory-frame $z$ axis defined by the direction of the external electric field, $A_M$ and $Z_M$ are the nucleon
and the proton number of the heavy nucleus in the molecule $M$, respectively.
The fit results are visualized in Fig.~\ref{fig:fitdeCS}. Apart from the individual constraints from the paramagnetic systems ThO,
HfF$^+$, YbF, Tl, we show the one from Hg, as well as the combinations of only the paramagnetic constraints and the global fit to all
systems. The fit
%
\begin{table}[th]
\centering{
\begin{tabular}{c c c c c c c c}\hline\hline
Molecule  & $\frac{E_{\rm eff}}{(\rm GV/cm)}$  & $\frac{W_S\footnote{Note the existence of different conventions in the
literature; for instance, the coefficient $W_S$ used here is called $W_{T,P}$ in Ref.~\cite{2017JChPh.147b1101S}, while $W_S$ in that
reference denotes the product $A/Z W_{T,P}$ appearing in Eq.~\eqref{eq::omegaM0}.}}{\rm kHz}$ & $\Omega$ & $\langle \hat n\cdot \hat
z\rangle$ & $\frac{\alpha_{d_e}^M}{\rm mrad/s/(10^{-27}e\,cm)}$ & $\frac{\alpha_{C_S}^M}{10^{7}\rm mrad/s}$   & Refs.\\\hline
HfF$^+$ & -23.0(0.9) & 20.4(0.8) & 1  & 1 & $34.9\pm1.4$ & $32.0\pm1.3$ &
\cite{2017JChPh.147b1101S,2017PhRvA..96d0502F,Cairncross:2017fip}\\
ThO  & -79.4(3.2)  & 112.1(4.5)  & 1  & 1 & $120.6\pm4.9$ & $181.6\pm7.3$ &
\cite{Denis-Fleig_ThO_JCP2016,2016JChPh.145u4301S,ACME_ThO_eEDM_science2014,Baron:2016obh}\\
YbF  & 23.1(1.8)  & -40.5(3.2)  & 1/2  & 0.558 & $-19.6\pm1.5$& $-17.6\pm2.0$ & 
\cite{Abe_YbF_PRA2014,Sunaga_YbF_PRA2016,hudson_hinds_YbF2011,Kara:2012ay}\\\hline
Atom & \multicolumn{4}{c}{$\alpha_{d_e}^A$} & \multicolumn{2}{c}{$\frac{\alpha_{C_S}^A}{10^{-20}e\,{\rm cm}}$} & Refs.\\\hline
Fr & \multicolumn{4}{c}{$885\pm35$} & \multicolumn{2}{c}{$1090\pm17$} &
\cite{2013PhRvA..88d2507R,doi:10.1021/jp904020s,2017PhRvA..95b2507S}\\
Tl & \multicolumn{4}{c}{$-573\pm20$\footnote{For discussions regarding this value, see also
Refs.~\cite{Nat_PRL_EDM_Tl,2012arXiv1202.5402N}. Note that the global fit is not affected by this discussion.}} & \multicolumn{2}{c}{$-700\pm 35$\footnote{See also Ref.~
\cite{Chaudhuri:2008xx}}} &
\cite{porsev_safronova_kozlov_tl2012,dzuba_flambaum_Cs_Tl_2009}\\
Cs & \multicolumn{4}{c}{$120\pm3$} & \multicolumn{2}{c}{$78\pm2$} &
\cite{2013PhRvA..88d2507R,dzuba_flambaum_Cs_Tl_2009,Chaudhuri:2008xx,Nat_PRL_EDM_Rb_Cs}\\
Rb & \multicolumn{4}{c}{$25.7\pm0.8$} & \multicolumn{2}{c}{$11.0\pm0.2$} & \cite{2013PhRvA..88d2507R,Nat_PRL_EDM_Rb_Cs}\\\hline
Hg & \multicolumn{4}{c}{$0.012\pm0.012$} & \multicolumn{2}{c}{$-0.028\pm0.006$} & \cite{mar-pendrill-oester_PS1987}, this work\\
\hline\hline
\end{tabular}
\caption{\label{tab::fitcoefficients} Relevant  information regarding the systems under consideration. $\alpha_{d_e,C_S}^A$ are
defined in analogy to Eq.~\eqref{eq::omegaM} as $d_A=\alpha_{d_e}d_e+\alpha_{C_S}C_S$.}} 
\end{table}
to only paramagnetic systems is massively improved by the HfF$^+$ measurement: before this measurement it extended essentially over the
whole green area. Our result for Mercury is seen to additionally improve the fit, reducing the
model-independent limits for both quantities significantly. This is due to the
constraint being essentially orthogonal to those from the paramagnetic systems: we obtain for the
paramagnetic systems a range $\alpha_{C_S}^{M,A}/\alpha_{d_e}^{M,A}\in [0.4,1.5]\times 10^{-20}e~{\rm cm}$, while for Mercury we
obtain conservatively $\alpha_{C_S}^{\rm Hg}/\alpha_{d_e}^{\rm Hg}<-0.9\times 10^{-20}e~{\rm cm}$. The latter ratio will
be more precisely determined once the coefficient for the electron EDM in Hg is known better, which is work in
progress; here we assumed an uncertainty of $100\%$, given the unreliable estimate. This improvement will also improve the
determination of $d_e$ and $C_S$.
In Table~\ref{tab::fitresults} we give the numerical results of both fits (global and paramagnetic only), including the effective
correlations between the results for $d_e$ and $C_S$, as well as the corresponding upper limits. While the individual constraints from
Hg are weaker than those extracted from ThO and HfF$^+$, its inclusion in the global fit results in model-independent limits about
a factor of two stronger than those from the paramagnetic systems alone.

\begin{table}[th]
\begin{tabular}{c c c c}\hline\hline
Fit  & $d_e/10^{-28}e$~cm  & $C_S/10^{-8}$  & Correlation\\\hline
global (w/ Hg)  & $\begin{array}{c}1.1\pm1.7\\|d_e|\leq 3.8\end{array}$  & 
$\begin{array}{c}-0.6\pm1.2\\|C_S|\leq 2.7\end{array}$  & $\begin{array}{c}-96\%\\\phantom{x}\end{array}$\\\hline
param. only (w/o Hg)  & $\begin{array}{c}-0.9\pm3.2\\|d_e|\leq 6.4\end{array}$  & 
$\begin{array}{c}0.8\pm2.4\\|C_S|\leq 4.9\end{array}$  & $\begin{array}{c}-99\%\\\phantom{x}\end{array}$\\\hline\hline
\end{tabular}
\caption{\label{tab::fitresults} Fit results for the global fit, using our result for Hg, and the fit using only the results from
paramagnetic systems. The former yields limits about a factor of two stronger than the latter.}
\end{table}

\section{Conclusions and Outlook}
We performed global fits to the available data constraining the electron EDM and the S-PS-ne nucleon-electron coupling entering
heavy atoms and molecules, using up-to-date calculations of the atomic and molecular structures. The inclusion of the recent result on
HfF${}^+$ ions improves drastically the fit to paramagnetic systems, only. As pointed out in
Ref.~\cite{MJung_robustlimit2013}, diamagnetic systems can be used to improve this fit additionally; while the corresponding
contributions are heavily suppressed in this case, diamagnetic systems have the advantage of constraining in some cases
combinations orthogonal to those accessible in paramagnetic systems. As an illustration we performed the first direct calculation of
the coefficient of the S-PS-ne coupling in Mercury, including the effect of electron correlations. In combination with the recently
improved experimental limit for this system we obtain limits on both the electron EDM and the S-PS-ne coupling of about a factor of two
stronger than from paramagnetic systems alone, see Table~\ref{tab::fitresults}.

Having a model-independent determination of both quantities determining the EDMs of paramagnetic systems in hand, we proceed to
evaluate the impact on on-going searches. The global fits imply non-trivial upper limits for every paramagnetic system that is not
effectively constraining the fits in Fig.~\ref{fig:fitdeCS}. These limits, given in Table~\ref{tab::Predictions}, indicate the
necessary precision for a given system to contribute significantly to the global fit or the fit to paramagnetic systems, only (given
in parentheses). A significant result above these limits would indicate an experimental problem. A measurement below the limit
from the fit to paramagnetic systems, but above the one from the global fit, could in principle also indicate the contrived
situation with a series of cancellations, described at the beginning section~\ref{SEC:PHENOMENOLOGY}.
%
\begin{table}[th]
\begin{tabular}{c c c c c}\hline\hline
Atom    & \multicolumn{2}{c}{Limits for $|d_A|/10^{-26}e$~cm}\\
& Inferred (this work) & Experimental\\\hline
Rb      & 0.7 (1.2) & $10^8$(1200)\cite{PhysRev.153.36,RbEDMthesis}\\
Cs      & 2.7 (4.2) & 1400 \cite{Murthy:1989zz}\\
Fr      & 13.0 (14.8) & ---\\\hline
Molecule& \multicolumn{2}{c}{Limits for $|\omega_M|/({\rm mrad/s})^*$}\\\hline
YbF     & 3.7 (5.6) & 27.8 \cite{hudson_hinds_2002,hudson_hinds_YbF2011,Kara:2012ay}\\
\hline\hline
\end{tabular}
\caption{\label{tab::Predictions} Model-independent limits for paramagnetic systems from our global fits;
the numbers in brackets correspond to the fit including paramagnetic systems, only. ${}^*$:~Assuming the same degree of
polarization as in the previous experiment.}
\end{table}

In the future, it is to be expected that measurements in paramagnetic systems alone will yield sufficiently precise results to limit or
determine the two contributions discussed here by themselves. In that case our calculations will serve to improve the model-independent
determination of hadronic contributions to diamagnetic EDMs in the context of a global fit extending over the whole set of P,T-odd
interactions.

\section{Acknowledgments}
 This research was supported by the DFG cluster of excellence ``Origin and Structure of the Universe''.
The authors are grateful to the Mainz Institute for Theoretical Physics (mitp) for its
hospitality and its partial support during the completion of this work. 
\bibliographystyle{unsrt}
\bibliography{all}
\clearpage

\end{document}